# NLRP3 monomer Functional Dynamics: from the Effects of Allosteric Binding to Implications for Drug Design

Emanuele Casali,[1] Stefano A. Serapian,[1] Eleonora Gianquinto,[2] Matteo Castelli,[1] Massimo Bertinaria,[2] Francesca Spyrakis,[2,*] Giorgio Colombo[1,*]

[1] Department of Chemistry, University of Pavia, Viale Taramelli 12, 27100 Pavia (Italy)

[2] Department of Drug Science and Technology, University of Turin, Via Pietro Giuria 9, 10125 Torino

* Authors to whom correspondence should be addressed:

G. Colombo; e-mail: g.colombo@unipv.it

F. Spyrakis; e-mail: francesca.spyrakis@unito.it

**ABSTRACT.**

The protein NLRP3 and its complexes are associated with an array of inflammatory pathologies, among which neurodegenerative, autoimmune, and metabolic diseases. Targeting the NLRP3 inflammasome represents a promising strategy for easing the symptoms of pathologic neuroinflammation. When the inflammasome is activated, NLRP3 undergoes a conformational change triggering the production of pro-inflammatory cytokines IL-1β and IL-18, as well as cell death by pyroptosis. NLRP3 nucleotide-binding and oligomerization (NACHT) domain plays a crucial role in this function by binding and hydrolysing ATP and is primarily responsible, together with conformational transitions involving the PYD domain, for the complex-assembly process. Allosteric ligands proved able to induce NLRP3 inhibition.

Herein, we examine the origins of allosteric inhibition of NLRP3. Through the use of molecular dynamics (MD) simulations and advanced analysis methods, we provide molecular-level insights into how allosteric binding affects protein structure and dynamics, remodelling of the conformational ensembles populated by the protein, with key reverberations on how NLRP3 is preorganized for assembly and ultimately function. The data are used to develop a Machine Learning model to define the protein as Active or Inactive, only based on the analysis of its internal dynamics. We propose this model as a novel tool to select allosteric ligands.

**KEYWORDS:** Protein-Protein Interactions; Molecular Dynamics; Allostery

**INTRODUCTION**

The SARS-CoV2 pandemic has had a disruptive impact on human health worldwide. While the disease was initially characterised by respiratory infections, a more variegated number of symptoms started to be recognized after the first few weeks into the pandemic. Of particular interest and concern were manifestations involving the brain and the neurological system: indeed, in many cases the COVID-19 infection has been associated with neurological disorders, especially involving brain tissues. These effects were soon associated with excess inflammation induced by SARS-CoV-2.[1] Moreover, an increased risk of developing neurological or psychiatric disorders in the six months following diagnosis was observed, with patients requiring hospitalisation and intensive care.[1],[2],[3]

One of the key molecular factors that shew to play a role in these detrimental mechanisms is the NLRP3 (Nucleotide-binding and oligomerization domain, Leucine-Rich repeat and Pyrin domain-containing 3) protein. NLRP3 is an intercellular sensor whose activation induces inflammasome formation and pyroptosis, which is also known to be involved in brain disorders and neuroinflammation.[4],[5],[6],[7]

It has been shown that the neuroimmune response and the disruptive passage of the virus through the blood brain barrier result in an increase of extracellular levels of ATP, with a concomitant activation of the ATP-gated ion channels of P2X7 receptors.[5] It is in this context that NLRP3 gets activated and assembled, with further increase of neuroinflammation.[5]

As a consequence, inhibition of inflammasome activation by blocking NLRP3 activation can provide novel direct opportunities to treat severe neurological manifestations of SARS-CoV-2. Furthermore, given its central role, the development of drugs directed against NLRP3 can potentially represent a general approach to the treatment of neuroinflammation.

During the last years, chemical scaffolds, in particular the di-substituted sulfonylurea typical of glyburide,[8] demonstrated their capability of inhibiting NLRP3. MCC950 (also known as CRID3) and NP3-146 are still the most active compounds, but their usage in clinics has been hampered by a certain level of toxicity.[9]

The publication of experimental structures of NLRP3 with or without ligands provided clues about the possible mechanism of inhibition. In particular, the X-ray structure of NLRP3 in complex with NP3-146 (PDB code 7alv)[9a] and the cryoEM one complexed with G2394 (PDB code 8etr),[10] and with ADP in both cases, revealed the stabilisation of a closed/inactive form, which likely implies blocking the closed/inactive to open/active transition of the protein. Recently, the cryo-EM structure of a polymeric aggregate in presence and in absence of MCC further demonstrated the influence of the ligand, not only on the single NLRP3 conformation, but also on the mechanism of aggregation. Indeed, as expected, MCC stabilises the closed ADP-bound form, which can aggregate into the formation of a homodimer decamer of intertwined leucine-rich repeat (LRR) domains that assemble back-to-back as pentamers (PDB code 7pzc; **Figure 1**).[11] Differently, the presence of the ATP-derivative AGS and the absence of inhibitors led to the formation of a disk-shaped active NLRP3 oligomer. Here, the central NACHT domain of NLRP3 assumes an ATP-bound conformation, in which two of its subdomains rotate by ~85 ° relative to the ADP-bound inactive conformation and the N-terminal PYDs from all subunits form, together, a PYD filament that recruits ASC PYD to elicit downstream signalling (PDB code 8ej4).[12]

This information could hold great potential in inspiring the design of new drugs: to meet this goal, it is crucial to complement structural data with approaches that permit to unveil the impact of the ligand (MCC in this case) on the functionally-oriented dynamics of NLRP3.

Here, we report a general comparative scheme based on the use of extensive Molecular Dynamics (MD) simulations to highlight the common and differential traits of NLRP3 dynamics in the inhibitor-bound and inhibitor-free states. In this framework, we developed a mechanistic model that reconnects the inhibitor induced effects on the protein's microscopic dynamics to the observed inactivation of NLRP3 by also proposing a new and efficient Machine Learning (ML) protocol for the analysis of protein dynamics data, to discern the dynamic signature differentiating active vs. inactive/inhibited states. This would allow classifying designed ligands as hits (those that induce inactive protein dynamic traits) vs. non-hits (those that do not alter the dynamics of the protein active state). The knowledge generated herein is currently being used to guide the design of novel allosteric ligands targeting NLRP3, thus expanding the chemical space of NLRP3 inhibitors and the range of possible interventions.

RESULTS AND DISCUSSION

**Structural organisation of NLRP3.** From the structural point of view, NLRP3 is organised in three main different domains: the *N*-terminal Pyrine domain (PYD), the central NACHT domain designated to accommodate ADP and the C-terminal Leucine-Rich Repeat (LRR).[11] A flexible linker combines PYD with NACHT. Geyer and co-workers previously hypothesised a key role for the PYD domain in driving the decamer supramolecular assembly (See **Figure 1** and vide infra).[11]

A finer subdivision of domains helps define other functionally important regions (**Figure 1**). The fish-specific NACHT associated domain (FISNA), located between PYD and NACTH domains (R176-K202), is associated to the conformational switch occurring during NLRP3 activation.[12] Finally, the acidic loop (K687-D700) close to the FISNA domain is fundamental for establishing electrostatic interactions with the nearby LRR domain,[11] and for mediating the contact between the concave site of two contiguous LLRs domains during the supramolecular assembly.[11]

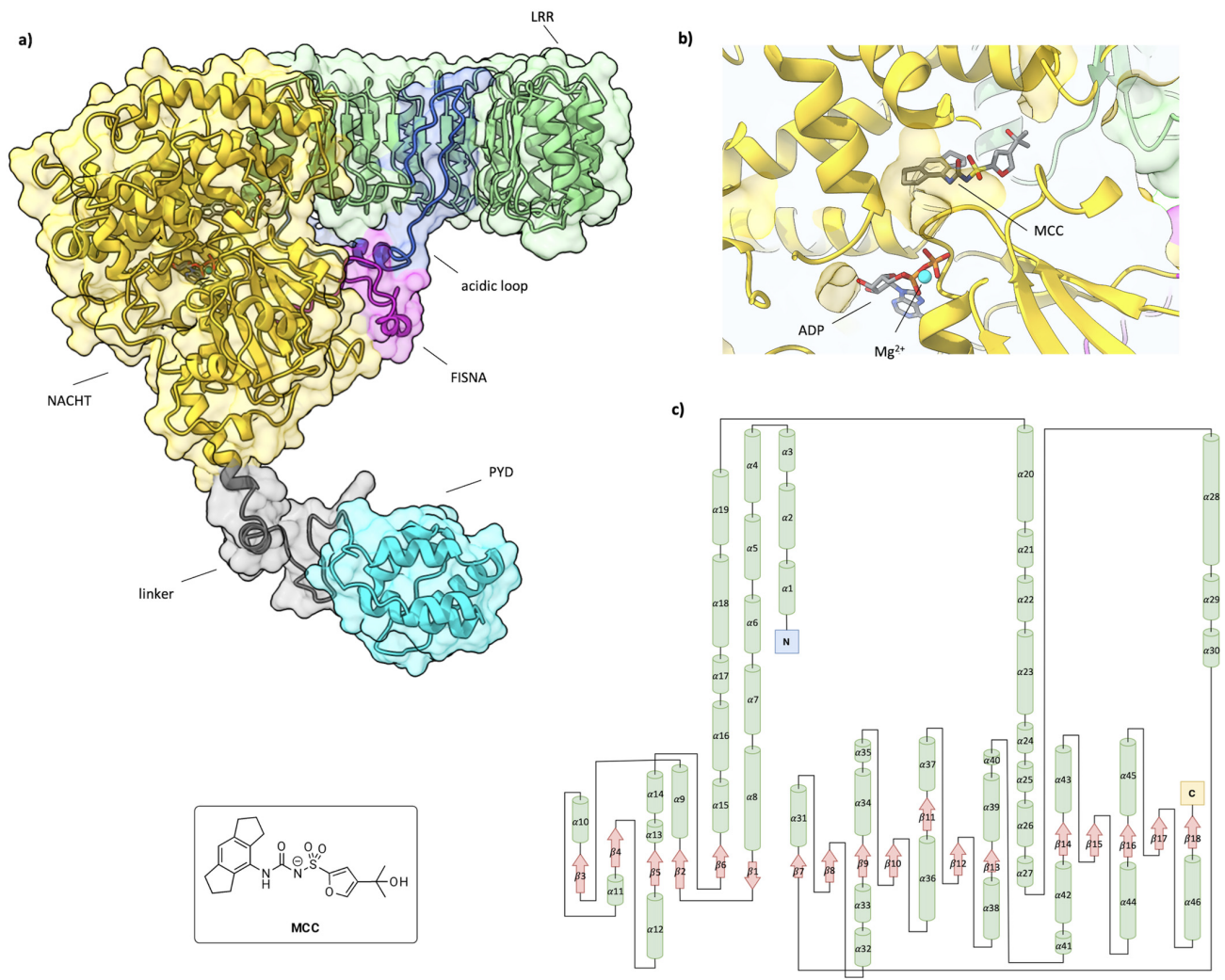

**Figure 1.** Cryo-EM structures of NLRP3 with MCC inhibitor: a) monomer with the three main domains highlighted, showing also the position of FISNA (magenta) and acidic loop (blue) regions; b) Zoom into the NACHT domain to reveal the MCC ligand structure and its binding pocket close to the ADP binding site ($Mg^{2+}$ ion coordinated); c) NLRP3 monomer sequence topology diagram and secondary structure numbering system.

The cryoEM structure of the monomer complexed with ADP and MCC (PDB ID 7pzc) was used as the starting point for extensive explicit solvent, all-atom MD simulations. Specifically, the latter were run on NLRP3 bound to ADP (simulation labelled **ADPstate**) and bound to both ADP and MCC (**MCCstate**). The total length of the simulations for each system was 4 μs.

**Characterization of the conformational space in the ADP and in the MCC states.** To garner a qualitative understanding of the ligand impact on NLRP3 dynamics, we first characterised large-scale conformational differences between ADP and MCC states, calculating the alpha carbon Root Mean Square Deviation (RMSD) for each system (**Figure 2**). It is immediately evident that both states are highly flexible (**Figure 2a,b**), with large-scale rearrangements substantially due to the PYD domain, which in the MCC state moves towards the LRR domain. Interestingly, distinct peaks in the histograms appear to indicate the presence of diverse structural ensembles for the two states. Representative structures of the various ensembles, taken from the largest bin in each peak, are reported as insets.

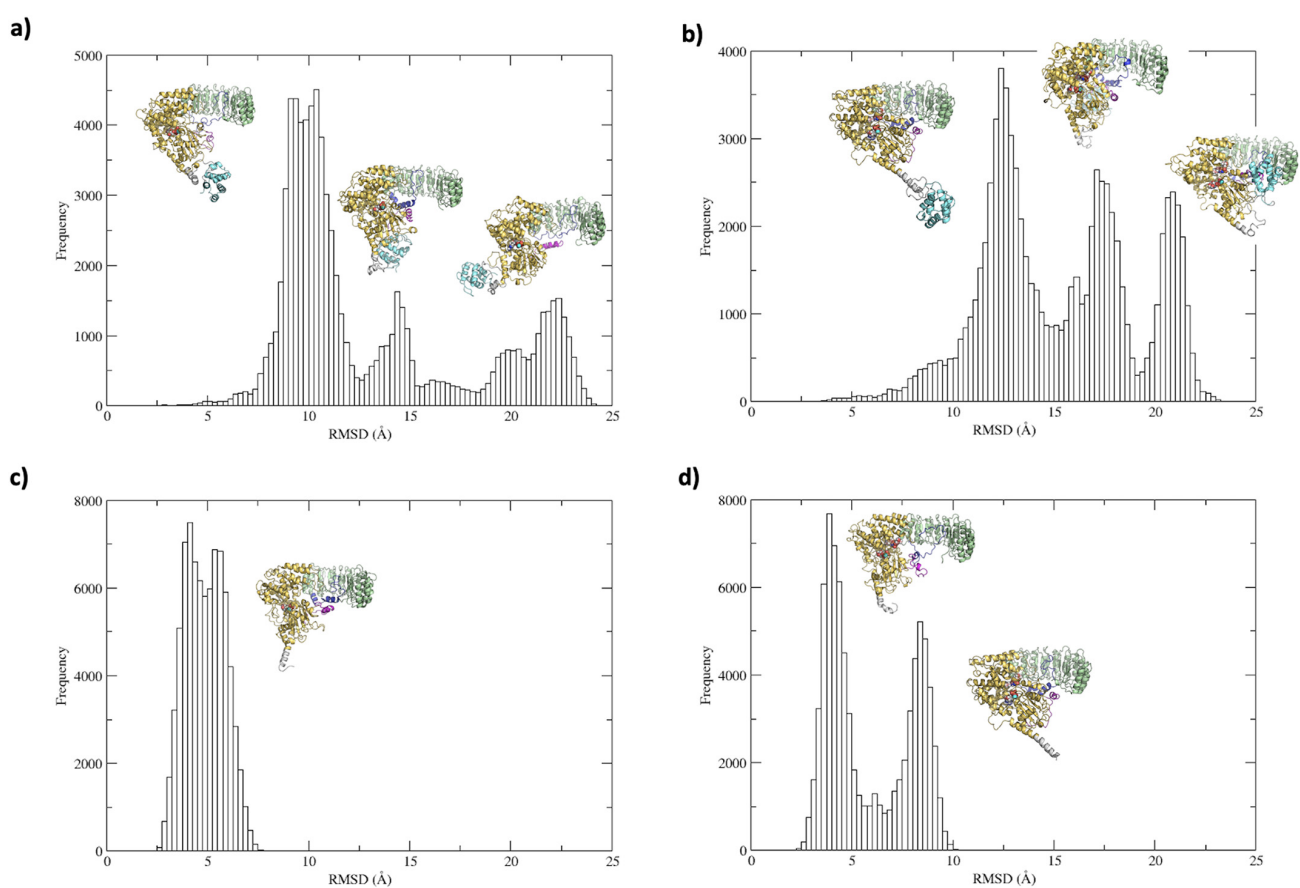

**Figure 2.** RMSD distributions and structural representatives of each ensemble: a) ADPstate using complete structure; b) MCCstate using complete structure; c) ADPstate after PYD removal; d) MCCstate after PYD removal.

To remove possible artefacts in the evaluation of conformational distributions due to the intrinsic flexibility of the PYD portion, we rerun the RMSD calculation considering only NACHT and LRR domains.

In the **ADPstate**, the RMSD distribution stabilises around a single peak (**Figure 2c**), while the presence of MCC (**Figure 2d**) appears to favour two distinct conformational ensembles. The first is similar to that assumed by the **ADPstate**, the second shows an opening motion between LRR and NACHT, which eventually creates the space for accommodating the PYD domain (**Figure 3**). The presence of MCC appears, thus, to induce a PYD rearrangement that make it inaccessible for establishing interactions with other monomers in the assembly of functional oligomers.[12]

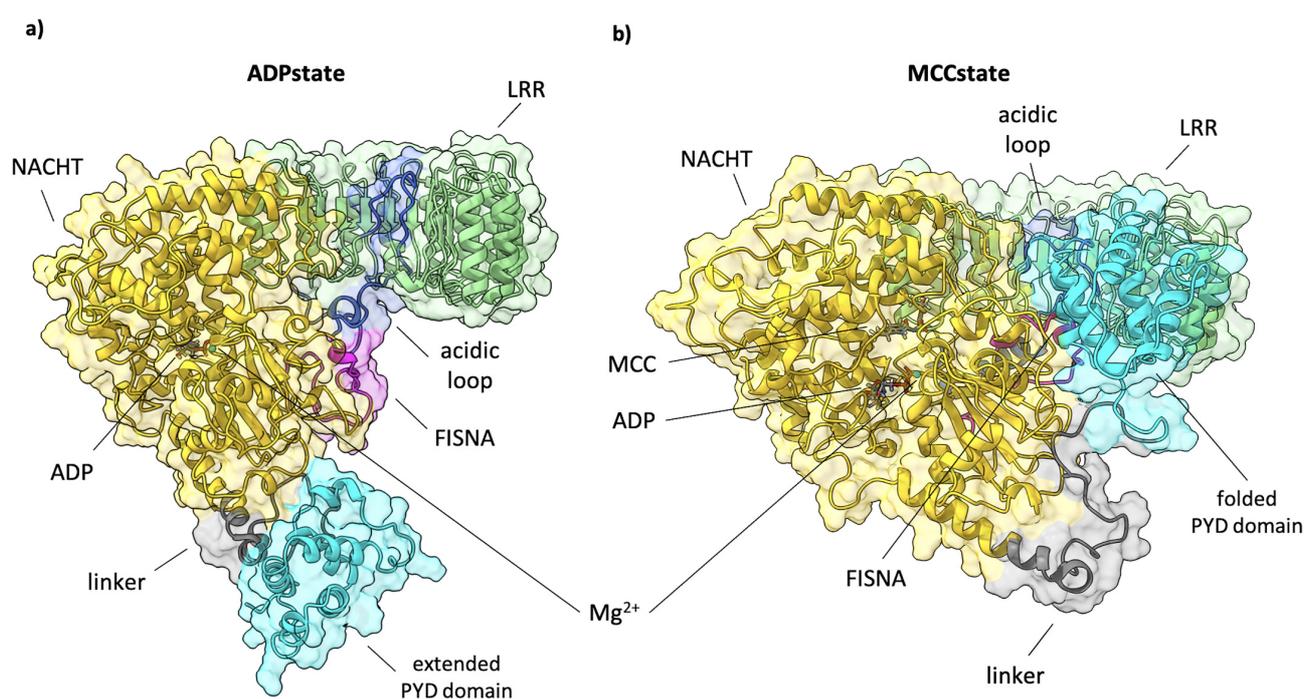

**Figure 3.** Representative structures of a NLRP3 monomer without (a) and with (b) the MCC ligand, showing PYD domain folding in the presence of the inhibitor.

The presence of MCC can also be related to a reorganisation of the linker portion (**Figure 3**; grey), which, while being in a stable α-helix in the absence of the inhibitor, partially unfolds and favours

PYD reorientation in the **MCCstate**. Similar observations were drawn when calculating the intramolecular distances between ADP and PYD, MCC and PYD, and MCC and ADP in the two (see Figure S1 in the Supplementary Information), considering for each substructure the centre of mass (COM).

Altogether, these first qualitative analyses show that the allosteric ligand significantly perturbs the overall dynamics of NLRP3 and favours conformational ensembles that are not poised to form the inter-monomer PYD-mediated interactions that eventually translate into the assembly of functional oligomers.

**Characterization of the internal dynamics of NLRP3 in ADP and MCC states.** To gain more insights into the impact of MCC on the protein internal dynamics in terms of short- and long-ranged perturbations, and ultimately, on its biological function, we computed the pairwise fluctuations of residue distances in all MD trajectories (see Materials and Methods). Intramolecular residue-pair distance fluctuations have previously been linked to the degree of coordination and allosteric communication between different protein substructures.[13],[14] In the present case, the goal is to investigate how changes in the structural dynamics of a protein on short time-scales can reverberate in the modulation of the (slower and possibly large-scale) motions that ultimately determine biological functions. The underlying hypothesis is that the comparative analysis of pair-distance fluctuations can reveal the (local as well as distal) disruption and reassembly of interactions that determine the structural deformations on the protein energy landscape underlying functionally oriented motions. Such mechanisms entail the selection of conformational ensembles and dynamic states of NLRP3 that are preorganized to form oligomers (the **ADPstate**) vs. states where the conformational organisation is unfavourable to oligomeric assembly.

The pairwise mean-square Distance Fluctuations (DF) values for the **ADPstate** and **MCCstate** are shown in the colour-coded matrices reported in **Figure 4.** The figure also reports the difference matrix between the two states, in which areas depicted in red denote a loss of coordination upon adding MCC, while darker (black) areas denote a loss of coordination upon removing MCC (for a full DF analysis in each replica, see Figures S4-6 in the Supporting Information).

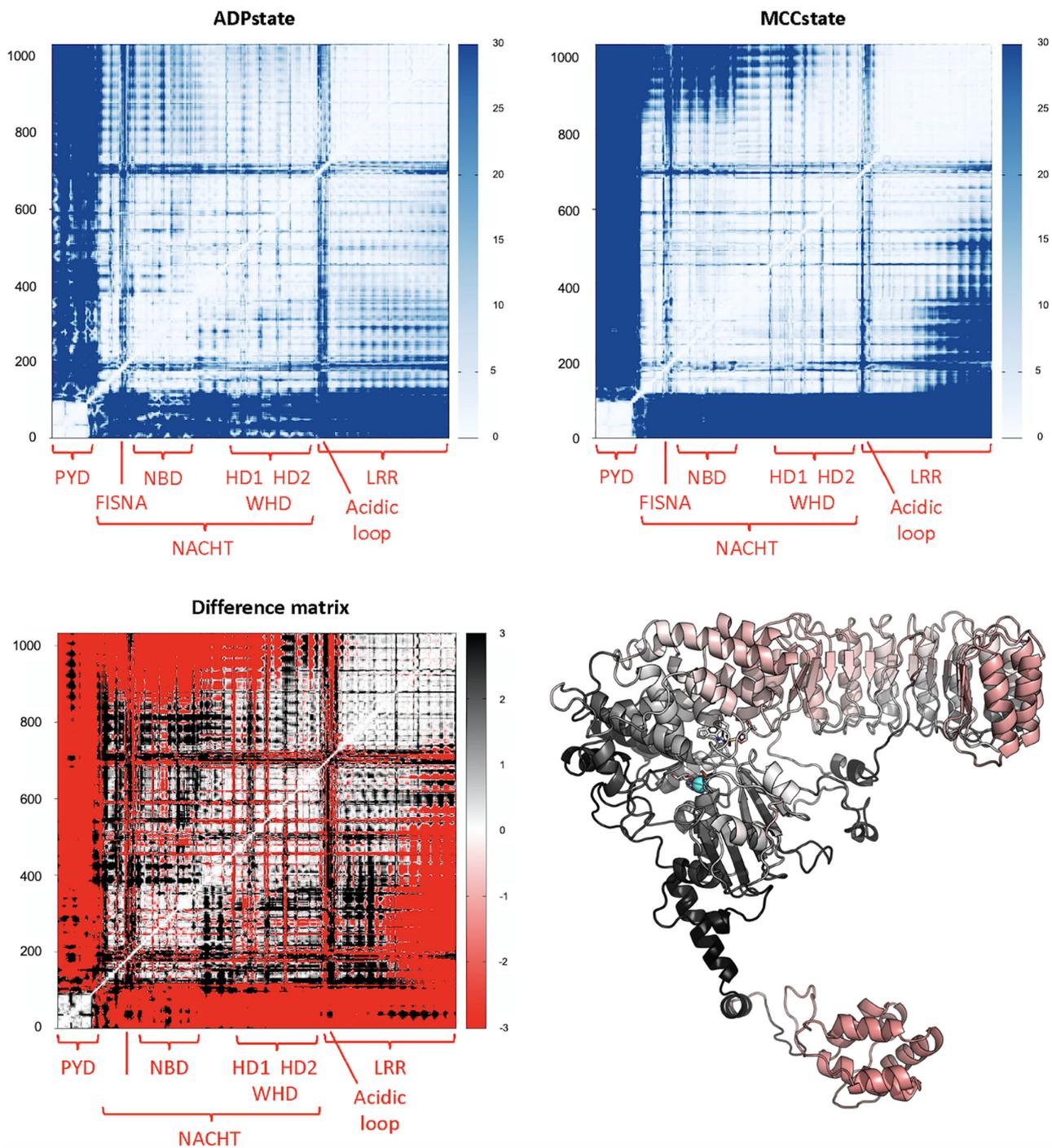

**Figure 4.** Distance fluctuation (DF) matrices and difference matrix. Structural representations of the difference matrix is also reported on the initial Cryo-EM structure, following the same color code:

redder areas denote a loss of coordination upon adding MCC; whereas blacker areas denote a loss of coordination upon removing MCC.

Overall, the matrices of both systems exhibit a block character, which is typical of multidomain proteins and reflects the alternation of regions with small inter-residue distance fluctuations (compact and internally coordinated domains), with regions of large ones (interdomain motions/loops). Differences, however, emerge examining the finer details of the matrices, which point to finely tuned MCC-dependent modulations of NLRP3 internal dynamics.

In the **ADPstate**, the PYD domain appears to be uncoordinated to the rest of the protein (i.e., high DF values; bluer areas). High coordination is only observed between PYD and certain regions of NACHT, namely NBD (where ADP is located), helical domain 1 (HD1), winged helix domain (WHD), and helical domain 2 (HD2). PYD and terminal LRR domains, on the other hand, display highly uncoordinated movement with respect to each other.

Focussing on the NACHT domain, the FISNA substructure appears here uncoordinated with respect to all the protein residues, while the NBD shows an extremely high level of coordination with the rest of the NACHT and the remainder of the protein. The LRR domain appears to be highly coordinated, both internally and at the inter-domain level, with the rest of the protein. Of particular interest is the acidic loop, known to mediate the molecular contacts between the concave sites of two opposing LRRs in the formation of the decamer complex:[11] during our simulations, in the absence of MCC, it showed uncoordinated motions with most of the protein, except with the closer residues of NACHT and LRR domains.

In the **MCCstate**, the PYD domain is characterised by a more diffuse network of intra-domain small fluctuation patterns, which is indicative of higher internal rigidity. At the same time, PYD loses coordination with all the remaining portions of the protein, in particular with those regions of the NACHT domain where low DF values were observed in the **ADPstate**.

Within NACHT, loss of coordination is observed between NBD and LRR domains. FISNA appears to be slightly more coordinated with the remaining NACHT residues, but still shows high DF values (poor coordination) with the LRR ones. The same situation can be found with the helical and winged domains, with an increase in fluctuations with LRR if compared to the **ADPstate**. The acidic loop is more coordinated with the internal residues of NACHT and LRR, but again uncoordinated with the terminal LRR ones. Finally, the LRR residues showed an increased level of coordination, given the lower DF values observed in this area.

Summarising, the results show a clear dependence of NLRP3 internal dynamics on the presence of MCC in the allosteric site. The allosteric ligand modifies the dynamic states of NLRP3, modulating the internal coordination patterns of substructures that are important to guide conformational transitions (primarily FISNA and acidic loop) and to sense the presence of the nucleotide in the binding site (the NBD). The perturbative effect of MCC reverberates on the internal dynamics and the intramolecular coordination patterns of the PYD domain, whose structural preorganization is important for functional oligomer assembly: interestingly, the PYD domain appears to lose the coordination with the rest of the protein in the presence of MCC. This, in turn, may favour a more disordered type of domain dynamics, which is consistent with the population of alternative conformations observed in the previous paragraphs.

**Ligand-dependent motions: fine mechanistic determinants and mechanical hotspots.** To reconnect the analysis of large-scale structural remodelling and of the fine-tuning of internal dynamics patterns with specific physico-chemical interactions between the ligands and the protein, we investigated the differences in the immediate vicinity of the binding sites of ADP and MCC. in the **ADPstate** and **MCCstate** (see also Tables S1-6 in the Supporting Information).

Residues making up the pockets that host ADP and MCC are displayed in **Figure 5.**

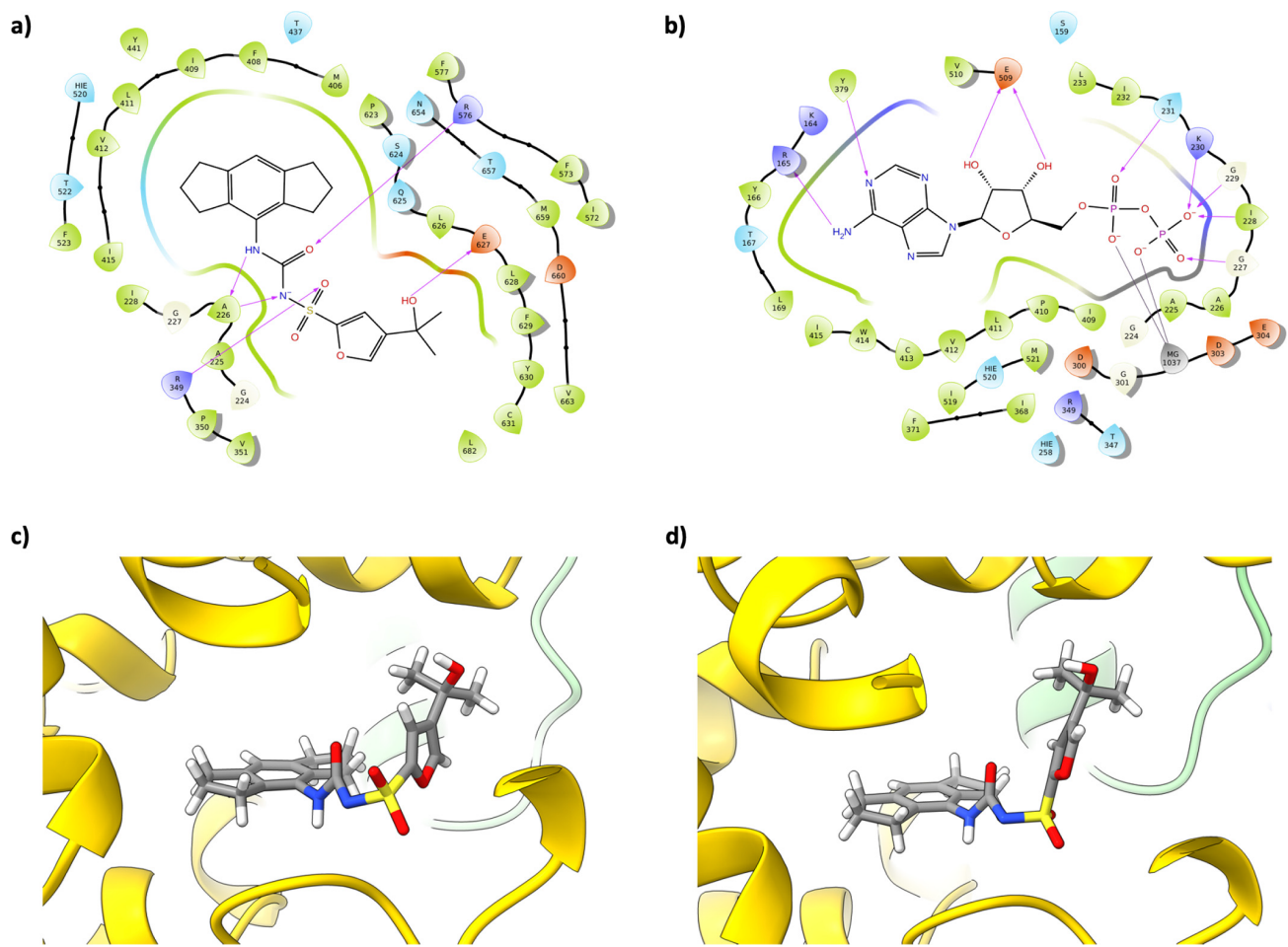

**Figure 5.** Ligand-protein interactions plot: a) MCC binding pocket and b) ADP binding pocket (purple arrows show the direct H-bonds between residue and ligand). We also show the two possible conformations of MCC through two contiguous bonds and around the dihedral N(urea)-S-C-O(furan): c) type-a (~90% of all frames of each replica) and d) type-b (~10% of all frames of each replica).

MCC and the protein establish a dynamic cross-talk: the ligand adapts to the protein by switching between two main conformations (**Figure 5a** and **5b**, see also Figure S10 in the Supporting Information for more details regarding the dihedral distribution), while the protein adapts to the ligand by dynamically displaying variable sets of residues for interaction. Overall, the OH group on the ligand establishes interactions with surrounding E627, P623, S624, and D660. At the same time, the furan and dimethyl moieties engage in hydrophobic interactions with the immediate surrounding. The large lipophilic portion of MCC, namely the indacene group, expectedly occupies a large hydrophobic region, where the large aromatic ring is flanked by M406, I409, L411, V412, T437, T438, T522 and M659. Stable interactions are formed by the sulfonamide group. The negatively charged nitrogen anchors MCC to the binding site together with the S=O function via H-bonding interactions with the backbone NH's of vicinal A225 and A226. Importantly, the negatively charged and highly polar sulfonamide oxygens strongly interact with the positively charged R349. This key interaction will be further discussed below (*vide infra*) as a key switch for allosteric control.

The contact analysis for the ADP binding site indicates the NBD pocket stably encloses the ligand for the whole simulation. Interestingly, the nucleotide binding site shares a number of residues that directly contact MCC, specifically A225, A226, L411, and R349. The latter, in the **ADPstate,** is stably sequestered by interactions with the nucleotide, generating a (local) state that is different from the ones observed for the **MCCstate**. The observation of shared contacts among the allosteric and orthosteric sites suggests the conformational perturbation encoded by MCC can be efficiently transferred, through this dynamic reorganisation of contact networks, to the NDB and the nearby FISNA activation domain. The time evolution profiles of the contacts are reported as Supporting Information in Figures S11-24.

**R349 as the allosteric switch that controls FISNA reorganisation and triggers the remodelling of NLRP3 dynamics: investigation of detailed interactions.**

Amongst the interactions between the protein and MCC, the one involving R349 appears to be one of the most interesting. While we were writing this paper, a publication appeared from Geyer and co-workers, showing that the interaction between ADP and sensor R349 is essential for the nucleotide hydrolysis in the active state of the protein.[15] R349 is instead sequestered by the sulfone group on MCC. This interaction is clearly observable in **Figure 6**.

To better characterise the mechanisms of R349 perturbations, a detailed monitoring of dihedral angles in the sidechain of R349 was carried out (see **Figure 6a**). Indeed, MCC sequesters R349, with notable shifts in the distribution of χ1, χ3 and χ4 compared to the **ADPstate**. (**Figure 6**; see also Figure S25 in the Supporting Information for more details).

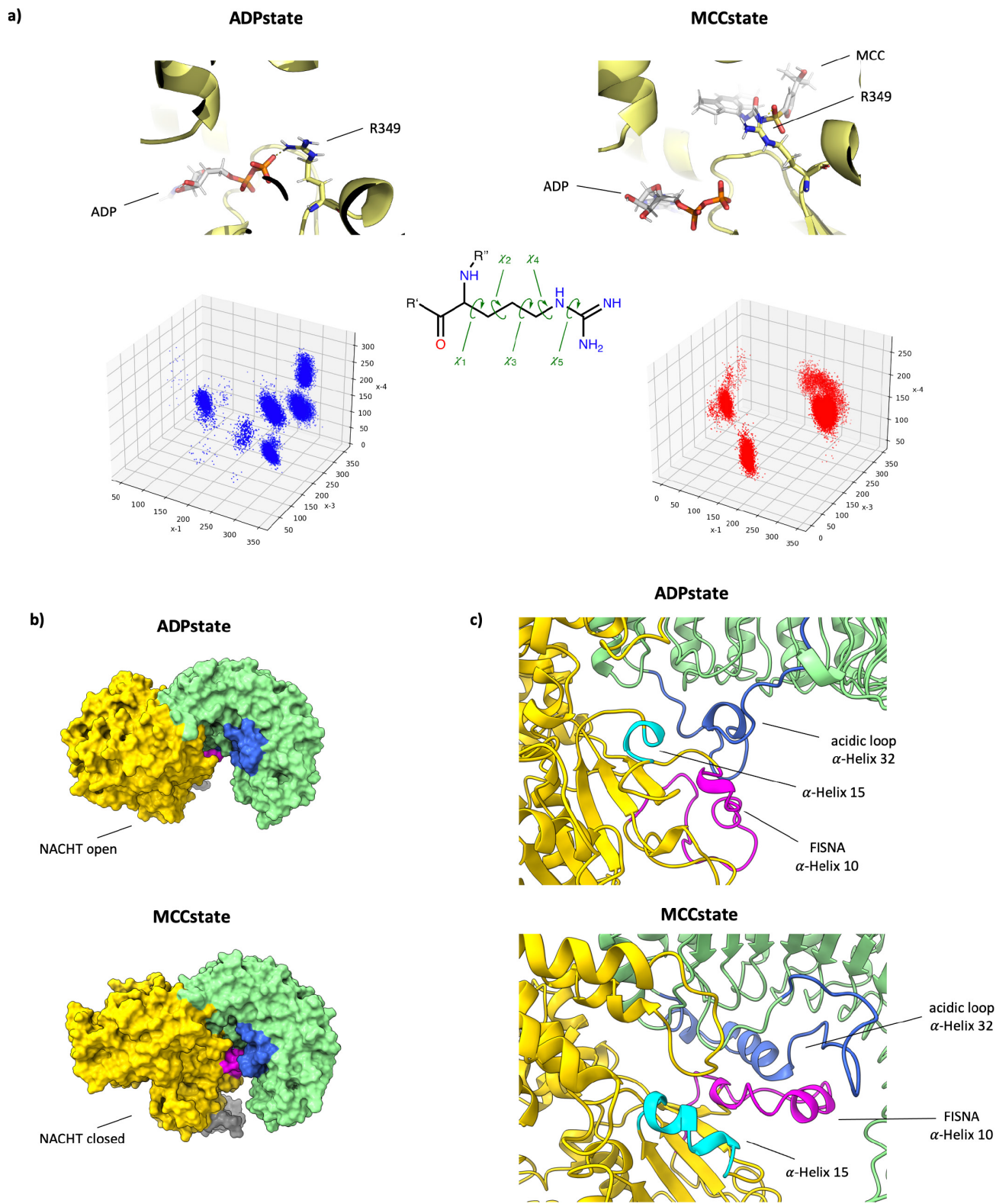

**Figure 6.** Allosteric modifications induced by R349 change in coordination: a) Different R349 orientations in clustered simulations for both ADPsate (left) and MCCstate (right) showing the difference in coordination of R349 which moves from ADP to MCC. Below, the five dihedrals of R349 side chain and the two spatial patterns observed during simulations: ADPstate (left) and MCCstate (right); b) Change in NACHT directionality towards (MCCstate) and away from (ADPstate) the LRR domain; c) Organization level of α-helix 10 (FISNA), 15 and 32 (acidic loop) in absence (ADPstate) and in presence (MCCstate) of the inhibitor.

The observed ligand-induced modification in the orientation of the R349 sidechain results in a reorganisation of the residues belonging to the central portion of the NACHT domain (See **Figure 6b**). In this context, R349 locks the ligand and drags along the subsequent α-helix (residues from R349 to L359). In doing so, it contributes to extending the length of the α-helix, which is limited to residues L353 to Q357 when MCC is absent. Loss of contacts by R349 with ADP in favour of new interactions with MCC emerges as one of the keys for the allosteric ligand-induced perturbation of NLRP3 functional dynamics.

These local modifications influence the vicinal FISNA activation group (residues from R176 to K202), consistent with the internal coordination analysis. In the absence of MCC the residues of this group remain mostly unorganised, with only one α-helix between Q183 and L186 (α-helix 10, see topology diagram in **Figure 1c** for more details on numbering system). Vice versa, when MCC is present α-helix 10 expands from R181 to A187, while the disordered loop between R176 and Q179 also folds up in α-helix, favouring the packing of the NACHT domain with LRR (See **Figure 6c**).

This finding is in agreement with recently published work by Wu and co-workers, who observed the crucial role played by FISNA central region in the protein structural reorganisation to stabilise the NACHT domain and access the supramolecular assembly of NLRP3 monomers (See **Figure 1**).[12]

Moreover, changes in the FISNA loop also impact on the vicinal α-helix (α-helix number 32) close to the acidic loop. Consistent with what previously observed, MCC promotes the helical organisation of residues spanning sequence V705 to S715; in the absence of the ligand, helix 32 spans a more limited number of residues, S709 to A713. This conformational modification also favours better packing between the NACHT and the LRR domains.[11]

**The role of water molecules in modulating MCC binding and effects.** Water molecules in binding sites are known to play an important role in favouring specific dynamics by stabilising specific interaction networks. In all the replicas, visual inspection permitted the identification of a subset of tightly bound water molecules in the vicinity of MCC (and ADP as well), in contrast to regions where local waters are in rapid exchange with the bulk. To put this observation on a more quantitative footing, we used the Waterswap approach on the representative structures of the most populated ensembles obtained from conformational clustering.[16],[17] In **Figure 7**, we report the network of stable water molecules and their H-bond networks. The data indicate that the waters identified as stable in the network approach the ligand from the pocket entrance, forming interactions that also entail R349 and MCC (see the Supporting Information for a full description of interactions). Such additional water-mediated interactions can provide further stabilisation to the bound state of MCC in the allosteric pocket of NLRP3.

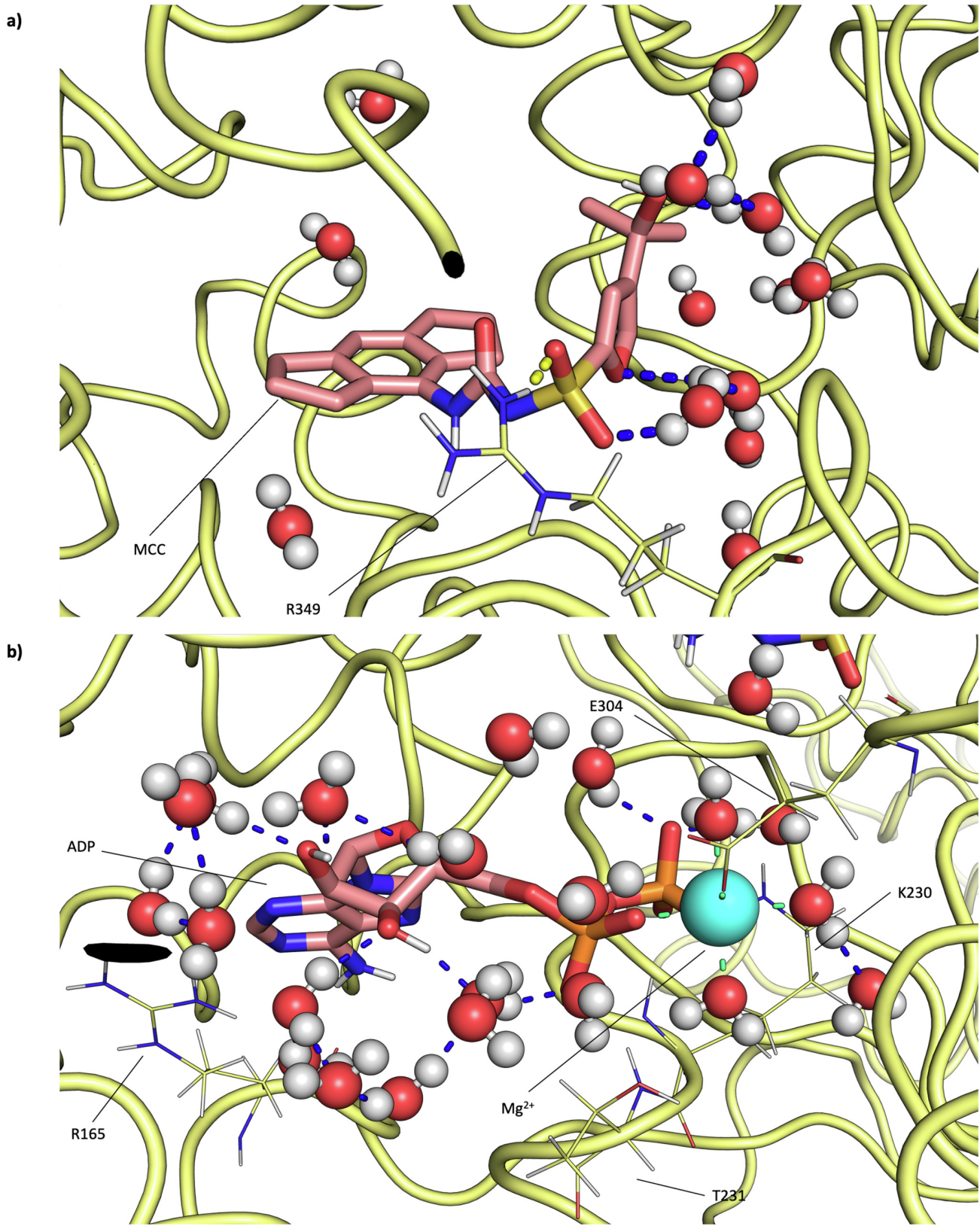

**Figure 7.** Water molecules involved in direct interactions with ADP and MCC ligands in the MCCstate: a) Water molecules and network of H-bond interactions with the MCC ligand (blue). In yellow the R349-MCC interaction is highlighted; b) Water molecules and network of H-bond

interactions with the ADP ligand (blue). In green the Mg-WAT interaction is highlighted. In both the cases, important residues are depicted.

In the case of ADP, we observed a complex interaction network involving the two hydroxyl groups of the ribose core and the surrounding water molecules. A second important region is located in the area where a water molecule acts as a bridge between the hydroxyl group of ADP and one of the oxygen atoms of the first phosphate group. Here T231 further stabilises the water molecules nearby. Finally, the terminal phosphate group coordinates two water molecules on the negatively charged oxygen. Interestingly, in the presence of MCC, K230 stabilises the terminal phosphate group, substituting R349 when this is involved in the interaction with the allosteric ligand.

Summarising, this detailed analysis of interactions in the **MCCstate** and **ADPstate** highlights how different factors cooperate in modifying the local dynamics around the two ligands in the two different states of the protein. MCC favours the switch of R349, further stabilised by a network of water molecules. The MCC-induced internal dynamics change leads to the extension of helices 10, 15 and 32, while also moving helix 15 (see **Figure 1c**), and increases the internal coordination of the FISNA region in the NACHT domain.

The combination of this detailed analysis with the characterization of internal dynamics reported above suggests that local effects brought about by MCC binding can diffuse throughout the structure, and modify the protein's overall conformational pre-organization necessary for aggregate into biologically functional complexes.

**The impact of MCC binding on Protein-Protein Interaction (PPI) Surfaces**

Ligand-binding can modify a protein's energy landscape and overall dynamics, remodelling surfaces that are distal to the actual ligand site. In the case of NLRP3, in particular, MCC binding can expectedly modify the (pre)organisation of interaction surfaces necessary to guide the assembly of functional supramolecular complexes. To investigate the impact of MCC on PPI surface modification, we predicted putative PPI interfaces in the absence and presence of MCC, using the Matrix of Low Coupling Energies (MLCE) method.[18]

Briefly speaking, MLCE is based on the analysis of the pair-interaction energies of all protein residues. For a protein of $N$ residues, it computes an $N \times N$ symmetric interaction matrix $M_{ij}$, where the matrix elements are the nonbonded part of the potential (van der Waals, electrostatic interactions, solvent effects). Eigenvalue decomposition of $M_{ij}$ highlights the regions of strongest and weakest couplings: the fragments that are on the surface, contiguous in space and weakly coupled to the protein core, define the potential interaction regions. In other words, putative interacting patches are assumed to be characterised by low stabilisation in their original setting and to be prone to be stabilised by a second partner. This framework is reminiscent of methods aimed to identify frustrated intramolecular interactions.[19]

We applied MLCE to the representative structures of the most populated conformational clusters extracted from the **ADPstate** and **MCCstate** simulations.

In the **ADPstate**, the top and front regions of the NACHT domain, highlighted in violet and red in **Figure 8a,** are predicted as possible interaction regions. This result is consistent with the recently published supramolecular, disk-shaped structure of polymeric NLRP3 (PDB ID 8ej4),[12] where the most relevant monomer-monomer interface involves exactly the two substructures of the NACHT domain here predicted (i.e. interface A in **Figure 8b**).

MLCE also defines the concave (internal) surface of LRR as a putative interaction region. Interestingly, in the above-mentioned supramolecular structure, this substructure engages in an interaction with the centrosomal kinase NEK7, which helps keep the structure open and extended.

In the **MCCstate**, a first potential PPI interface is located on the external surface of the LRR domain, in agreement the assembly of the inactive decamer (PDB ID 7pzc),[11] where two distinct LRRs from two consecutive monomers establish contacts through the predicted interface (**Figure 8c,** orange surface and Figure 8d, interface A). This result is also consistent with observations from Geyer and co-workers who implied this interaction as an important stabilising factor in the decamer assembly.[11] The second PPI interface here predicted is located on (the external portion of) the NACHT domain, specifically involving the helices in the HD2 subdomain (**Figure 8c**, green surface). Importantly, in the inactive decamer, this region is involved in interactions with the terminal portion of the LRR domain and the third and fourth helix of HD2 subdomain of two consecutive monomers (**Figure 8d**, interface B).[11] Finally, MLCE returned a prediction of PPIs on the PYD domain (**Figure 8c**, magenta surface), which is also corroborated by the PYD-PYD interactions observed experimentally in the assembly of the inactive decamer (see **Figure 8e**).

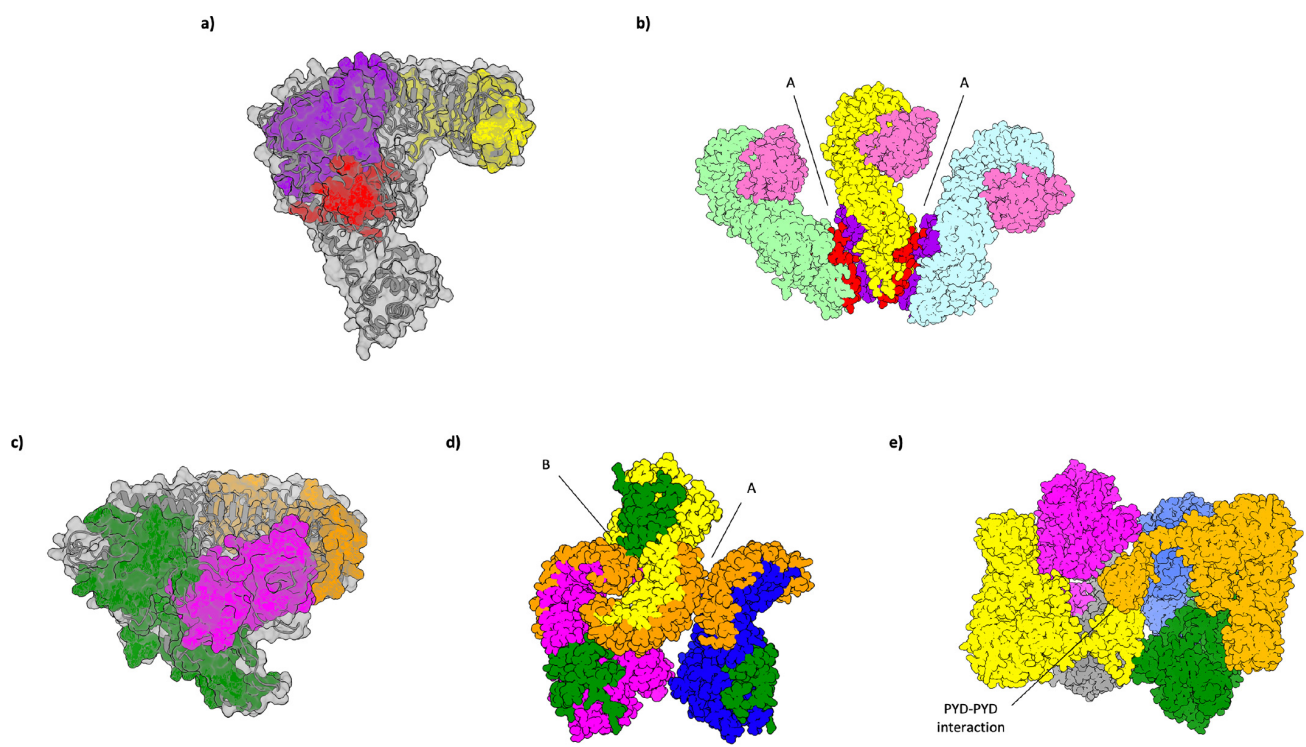

**Figure 8.** MLCE study on the (a,b) ADPstate and (c,d) MCCstate: a) ADPstate monomer with the three PPIs surfaces predicted (red and violet surfaces represent the possible interaction regions); b)

Section of three monomers in disk (ADPstate) highlighting the main predicted interface A which corresponds to the contact between red and violet surfaces belonging to two contiguous monomers; c) MCCstate monomer with the three PPIs surfaces predicted (purple, green and organge surfaces represent the possible interaction regions); d) Section of three monomers in the decameric double-disk (MCCstate) highlighting the two main predicted interfaces A (between orange areas belonging to two intercalated monomers) and B (between the terminal orange surface and the green of the contiguous monomer). The purple area in c) finds confirmation in the PYD-PYD contact observed in the centre of the double-disk decamer assembly as shown in d) which displays the PYD-PYD interaction at the center of decamer.[11]

Summarising, the energy-based prediction of potential PPI interfaces indicates that different likely interacting substructures may emerge and be presented for partner-binding as a function of the ligand-state of the protein. In our model, the presence of MCC determines a rearrangement consistent with a model where monomers are preorganized to assemble via interactions observed in the recently described decamer form: this is achieved via the specific interaction interfaces, that present distinct structural and physico-chemical profiles from the ones observed in the **ADPstate**.

**Proceeding towards New Drug Selection: MD-based Machine Learning Classification of NLRP3 Active vs. Inactive States.**

The results reported above provide a detailed view of the mechanisms of ligand-based regulation of NLRP3 functions, and specifically of the impact of the presence of only ADP or ADP in combination with MCC on potential interacting surfaces. The proposed model is, in fact, able to differentiate functionally inactive dynamic states induced by the presence of MCC, from biologically active ensembles. This would allow the automatic classification of protein ligand-induced states as active or inactive via a simple analysis of the DF matrix images originating from different trajectories. To meet this challenge, we further developed and tested a Machine Learning approach that, starting from the analysis of images corresponding to Distance Fluctuation (DF) matrices (see above), which give a compact account of the main internal dynamic traits observed in MD simulations, can classify the protein as active or inactive. Indeed, DF images capture the overall state of the protein, since small modifications within the protein structural organisation (such as ligand binding or even mutation) can impact on the overall coordination propensity.[13],[14]

To classify the protein state as Active or Inactive, we applied Convolutional Neural Networks (CNN). Specifically, we used the VGG19 classification algorithm,[20] directly available from Tensorflow (TF), and showing an optimal compromise between computational cost and accuracy.[20] We introduced small modifications to increase the dimensions of the layers according to the pixel number of the input images (See **Figure 9**).

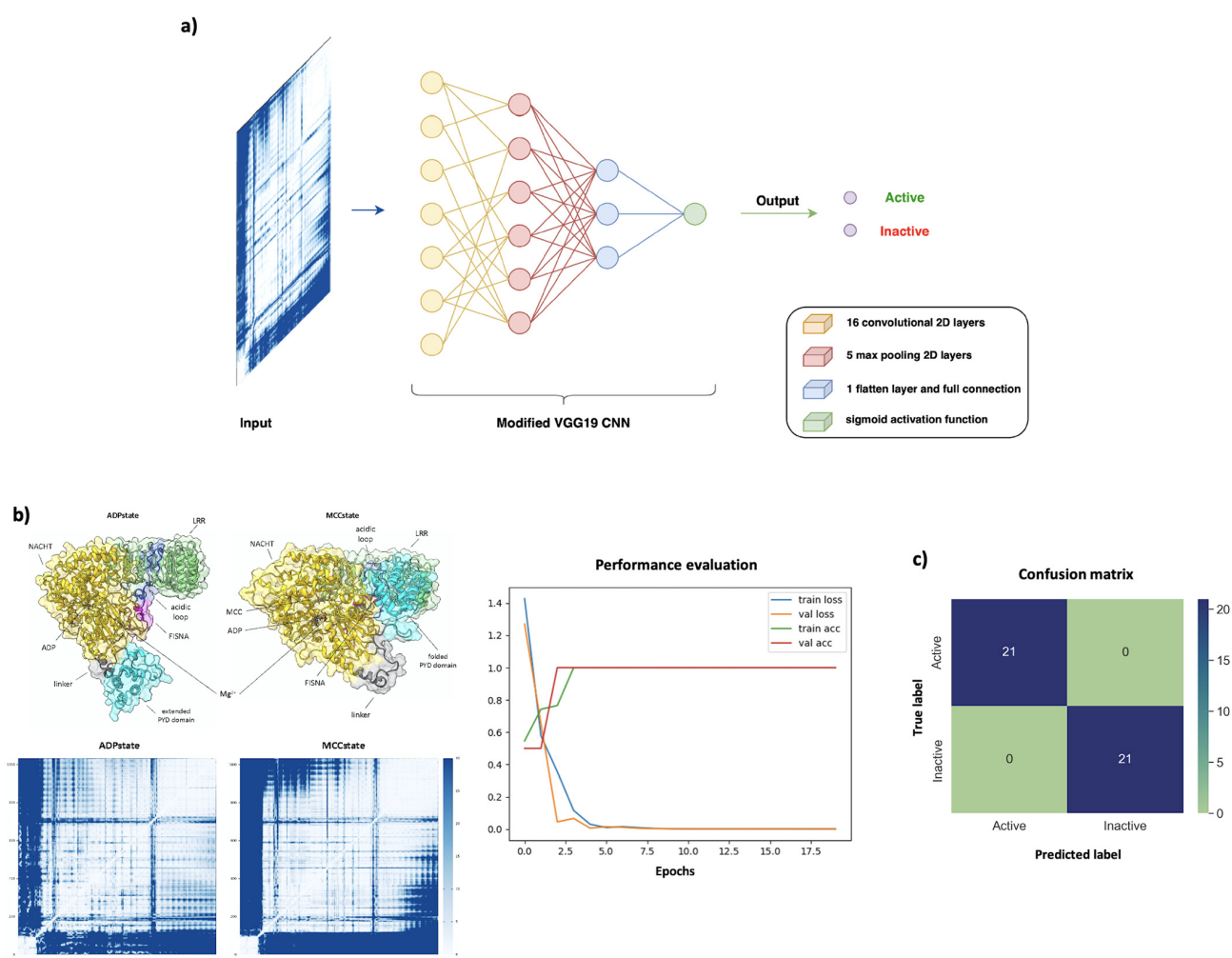

**Figure 9.** a) The architecture of modified VGG19 mode; b) different states of the protein involved in the study with the DF matrices (left) and performance evaluation of training and validation of the CNN ML-model (right); c) confusion matrix for the test data set using the trained ML-model.

The full detail of the procedure is reported in Materials and Methods, while the architecture and summary of the model is reported in the Supporting Information.

We firstly selected DF images from the last equilibrated 250ns of each of the four replicas for both the **ADPstate** and **MCCstate** of the protein, to generate images for model training. To prepare the datasets we extracted a DF image each 10ns. Considering the number of replicas for each variant, we ended up with a total of 208 images, on which we operated a manual random separation between test (20%), train (64%) and validation (16%) sets.

Our dataset was composed as follows:

• Training set: 132 DF matrix images (66 **ADPstate** and 66 **MCCstate**)

• Validation set: 34 DF matrix images (17 **ADPstate** and 17 **MCCstate**)

• Test set: 42 DF matrix images (21 **ADPstate** and 21 **MCCstate**)

The imported images were firstly rescaled to the dimension of the VGG19 layers and were normalized according with the standard pixel values, ranging from 0 to 255. We next defined the classification mode: since we wanted to simply perform a binary classification between Active (Corresponding to the **ADPstate**), and Inactive (Corresponding to the **MCCstate**) states of the protein, we set the classification mode to 'binary' (class_mode) and the number of samples propagated through the network was set to 32 (batch_size, see Python Script in the Supporting Information). To provide a measure for goodness of the method we used ImageNet weights as widely recognized to be a standard for images classification problems.[21]

The last layer of our VGG19 modified model provides the prediction output. This was achieved through a single layer smoothed by the sigmoid activation function σ(z):

$$\sigma(z) = \frac{1}{1 + e^z}$$

Given its existence only between 0 and 1, it constitutes the natural choice for binary classification problems. We compiled the model by using a binary cross-entropy loss function and using the Adam algorithm for the stochastic optimization.[22] Lastly, in order to avoid model overfitting, we introduced an early stopping monitor, which stops training the model if the validation loss starts increasing during five consecutive epochs. We trained our model for 20 epochs using the datasets and we obtained complete training and validation in 21 seconds, with an average of 147 ms/step. This result is extremely promising and is mainly due to the TF parallelization of using GPU.

The performance of our method on training and validation sets are shown in **Figure 9b**. In both cases we obtained a good mutual fit and convergence between validation and training accuracies as well as losses. It can be clearly observed how even small differences between active and inactive states can be immediately captured by the model and DF image analysis, and traced back to the respective ligand-states (**ADPstates** vs. **MCCstates**).

By evaluating our trained model on the test set, we got 100% of accuracy, taking less than 1s to scan all the 42 DF test images. This result is corroborated by the confusion matrix, which perfectly showed that amongst 42 total images, 21 test entries were correctly classified as Active while the other 21 as Inactive (**Figure 9c**).

Overall, the combination of MD simulations and ML image recognition proves able to correctly classify different states of a protein, based only on the analysis of its internal dynamics.

The potential of this classification method can allow us to predict the effect of other types of small molecule effectors binding in the MCC site, even with completely different structures, whose functional role is not known a priori. Given the difficulties in generating SARs for allosteric drug candidates, the use of dynamical descriptors and ML can help prioritise compounds with desired functional properties. This result is presented as a verifiable experimental hypothesis and is currently being used for the design/screening of a novel generation of NLRP3 inhibitors.

**CONCLUSIONS**

Herein we examined the origin of allosteric inhibition exerted on NLRP3 by MCC. Through the use of extensive MD simulations and comparative analyses, we have provided molecular-level insights into the mechanisms through which the allosteric effector can affect protein structure, dynamics, and functions. From MD simulations we observed the selection of significantly different conformational ensembles for the **ADPstate** vs. the **MCCstate**, an increased overall rigidity for the allosterically bound state, and distinct dynamic profiles in several key regions. We identified specific interaction and hydration patterns for the orthosteric and allosteric binding sites in the two states: in particular a key residue, R349 is shown to act as a switch sensing the presence/absence of the allosteric ligand and triggering the remodelling of local interactions and secondary structures. The latter are instrumental to transmit the molecular perturbation encoded by the inhibitor at distal regions. The presence/absence of the allosteric ligand also reverberates in the emergence of distinct intramolecular pair-energy patterns, which underlie the selection of specific substructures as possible PPI interfaces. Importantly, the presence of MCC is shown to induce a state for which the regions predicted as PPI interfaces optimally trace the ones observed experimentally in the supramolecular assembly of the inactive state of NLRP3.

Finally, our data are instrumental in developing a simple and direct Machine Learning model that permits to correctly classify the dynamic states of the protein as Active or Inactive, based only on the analysis of the internal dynamics of the protein and without the need for providing explicit information on the presence and nature of possible ligands. While classical physico-chemical approaches should be still employed to optimize a certain hit after the first selection via ML, our approach can be useful in screening candidates based on the effects they may have on protein dynamics. In particular, we suggest that a promising strategy would be to combine MD simulations of the protein in the presence of designed/screened molecules with the ML-classification model and select those ligands that induce Inactive states (according to the ML model) for further development and analysis. Thus, once these molecules have been identified, "focused" medicinal chemistry evolution studies could be designed to optimise their pharmacological profiles. Our laboratories are currently working on these further developments and results will be available soon.

## COMPUTATIONAL METHODS

### Parametrization of MCC ligand

MCC ligand preparation was performed by extracting the partial charges and the initial optimized coordinates from DFT calculations. Gaussian 16 program package was used to perform this step.[23] The optimization step was carried out using the B3LYP hybrid functional for DFT calculation.[24] We used 6-31+g(2d,p) basis set for all the atoms in the ligand.[25] Tight convergence criteria for the SCF cycles has been adopted (*SCF=tight*), while all the calculations were performed *in vacuo*. With the optimized structures, we performed single point calculations at Hartree-Fock level of theory, but by using the same basis set as before.[26] Specific options were included in this calculation step to directly obtain the charges according to the Merz-Singh-Kollman scheme and atom types to be submitted to the AmberTools parametrization [i.e., iop(6/33=2,6/41=10,6/42=17,6/50=1)].[27] This results in the preparation of the .gesp file that will be used in the next step of preparation of Amber submission file. The parameters for the ADP cofactor and $Mg^{2+}$ ion were directly used from the Amber parameter database provided by University of Manchester.[28]

### Preparation of protein

NLRP3 monomer was directly taken bound to the MCC ligand, as provided by the RCSB Protein Data Bank and deposited by Geyer *et al*. as a cryo-EM crystal structure (PDB ID 7pzc).[11] Maestro20 Schrödinger suite and AmberTools21 were used during the initial preparation of the protein, in order to complete missing portions (Maestro Prime tool) and generate atom types and coordinates in a compatible format with Amber software.[29],[30] AmberTools21 software suite was used for the preparation of files for MD simulations, by taking the prepared structures of both NLRP3 protein – with and without the MCC ligand. We used AMBER's *Tleap* tool to prepare Amber submission file: the well parametrized FF14SB force field was selected to treat the entire environment, while we adopted a cubic solvation cage with water molecules treated with the TIP3P model.[31],[32] One

Magnesium ion was added. Consistent with previous work, the ion was placed to be complexed by the two phosphate ions of the nucleotide this stabilising ADP in the pocket.[33] Sodium ions were randomly added in order to neutralize the charge of the system.

**Molecular Dynamics details**

Simulation of the monomer structure was performed using Amber software (version 20) for four independent replicas. To generate each replica, we performed two rounds of minimization, each comprising 300 steps. This is followed by a preproduction phase lasting 2.069 ns, during which we use the *Sander* molecular dynamics (MD) engine. Subsequently, we switch to the GPU-accelerated *pmemd.cuda* MD engine for the remaining stages of preproduction as well as the production phase, which lasts for 1μs.[29]

Before the production stage, each MD replica undergoes several pre-production steps, including minimization, solvent equilibration, system heating, and equilibration. The first two steps are performed using the *Sander* utility on the CPU, and then we switch to the GPU-accelerated *pmemd.cuda*.

The minimization step comprises two rounds of 300 steps each. In the first round, we use the steepest-descent algorithm for the first 10 steps and then switch to the conjugate gradient method for the remaining 290 steps. During the first round, we only minimize the backbone Hα, while we restrain all other atoms harmonically with a force constant $k$ of 5.0 kcal mol$^{-1}$ Å$^{-2}$. In the second round, we release all atoms, including solvent and ions, for unconstrained minimization.

During the solvent equilibration step, we perform simulations in the *NVT* ensemble for a duration of 9 ps using a time step of 1 fs. We use the Berendsen thermostat[34] to maintain the temperature of the system, with non-solvent atoms harmonically restrained with a force constant $k$ of 10 kcal mol$^{-1}$ Å$^{-2}$. The solvent molecules are given random velocities to match a temperature of 25 K. We then rapidly heat the system to 400 K over the first 3 ps (coupling time: 0.2 ps), maintain it at 400 K for 3 ps, and finally cool it back down to 25 K over the last 3 ps with a slower coupling time of 2.0 ps. For this step and all subsequent ones, we use an 8.0 Å cutoff to determine Lennard-Jones and Coulomb interactions, and the Particle Mesh Ewald method is used to compute Coulomb interactions beyond this limit. Unlike other stages, we do not apply SHAKE and SETTLE constraints during solvent equilibration, but we do apply them in all subsequent stages.[35],[36]

During the system heating stage, we increase the time step to 2 fs and maintain the *NVT* ensemble, but we use the Langevin thermostat to enforce the temperature instead of the Berendsen thermostat, and we use it for all subsequent stages.[37] We start with an initial collision frequency of 0.75 ps$^{-1}$ and heat the system from 25 K to 300 K over a period of 20 ps. During this process, all atoms except amino acids' Cα atoms are allowed to move freely, while the Cα atoms are positionally restrained with a force constant *k* of 5 kcal mol$^{-1}$ Å$^{-2}$.

During the equilibration step, we switch the ensemble to *NpT* (with a pressure of 1 bar) and use Berendsen's barostat with a coupling time of 1 ps. We simulate the system for an additional 2040 ps with a thermostat collision frequency lower than that used in the production stage (1 ps$^{-1}$). We gradually remove the restraints on Cα atoms during this process: for the first 20 ps, the force constant *k* is 3.75 kcal mol$^{-1}$ Å$^{-2}$, for the following 20 ps, it is 1.75 kcal mol$^{-1}$ Å$^{-2}$, while there are no restraints thereafter.

For the 1 μs production stage, we use the *NpT* ensemble at a temperature of 300 K and pressure of 1 bar by keeping a 2 fs time step. The calculation of Lennard-Jones and Coulomb interactions employs an 8.0 Å cutoff (and does so during preproduction too). For Coulomb interactions beyond this limit, we use the Particle Mesh Ewald method,[35] Lennard-Jones interactions are not computed. We restrain all bonds containing hydrogen using the *SETTLE* and *SHAKE* algorithms[36] (for bonds in water and non-water molecules, respectively). To maintain constant pressure, we utilize Berendsen's barostat with a relaxation time of 1 ps,[34] while the temperature is kept stable using Langevin's thermostat with a collision frequency of 1 ps$^{-1}$.[37]

**Trajectory analysis**

The trajectories obtained were analyzed using the *cpptraj* module in AmberTools21.[29] The resulting RMSD, RMSF, contacts analysis and temporal evolution, as well as dihedral and distance charts were represented using Xmgrace. See Supporting Information file for more details.

**Residue-pair distance fluctuations (DFs)**

We computed the matrix of distance fluctuations (DF) by using both the 4 μs metatrajectory available for each studied system (which was obtained by concatenating the MD replicas of each specific protein) and the metatrajectory performed on the last 250 nanoseconds of each replica, thus representing the most equilibrated portion of the production (see Supporting Information for more details). Each element of the matrix corresponds to the DF parameters.

$$DF_{ij} = \langle (d_{ij} - \langle d_{ij} \rangle)^2 \rangle$$

The DF matrix elements correspond to the DF parameters, which are defined as the time-averaged distance between the Cα atoms of amino acids *i* and *j* ($d_{ij}$). Unlike the covariance matrix, the DF matrix is invariant under molecule translations and rotations and does not depend on a specific reference structure for the protein.

For each point in the trajectory, DF was computed for every residue pair. This metric identifies residues that move in synchronization, indicating the presence of distinct coordination patterns and quasi-rigid domains motion in the protein under investigation. Specifically, amino acid pairs that belong to the same quasi-rigid domain or are closely coordinated display minor distance fluctuations, while pairs that are not coordinated exhibit larger fluctuations.

A useful method for comparing fluctuation matrices is to make the difference point by point, here reported in **Figure 4**. Then it is possible to subtract from a reference matrix (**ADPstate**) the other one (**MCCstate**) by performing the difference operation residue by residue.

**Water-MCC and Water-ADP interactions**

In order to perform the water-contact analysis both with MCC and ADP we used the Waterswap application implemented by Mulholland and co-workers in the Sire software suite.[38] It works by identifying the most important water molecules interacting with the ligand interested by swapping it with a water cluster along a reaction coordinate.[38] We thus performed the calculation by using *cpptraj* module in the AmberTools21.[29] The most recurrent clustered structure in the **MCCstate** simulations was obtained by aligning trajectory frames to the MCC ligand and screening them according with the residues around the ligand identified during the contact analysis. The hierarchical agglomerative algorithm approach was involved, while the minimum distance between clusters was set greater than 26.0Å as the end point of the calculation, by keeping as final result the most representative 6.0 clusters.

The most representative frame was selected as the one with the highest fraction of total frames in the cluster and then used during the Waterswap calculations, to identify within the output files the most important water molecule's directly interacting with MCC or ADP.

**MLCE analysis for PPIs prediction**

One of the direct computational strategies to predict protein-protein interfaces is the use of the Matrix of Low Coupling Energies (MLCE) energy decomposition method.[18]

MLCE starts by analyzing the pairwise interaction energies of all amino acids in a protein.[18] The method has been previously described and experimentally tested under various conditions by our group.[18] However, it is useful to provide the reader with a brief overview. MLCE operates in a multistep process, in which the first calculates the unbound part of the potential $E$ (van der Waals, electrostatic interactions, solvent effects) through an MM/GBSA calculation, obtaining, for a protein of $N$ residues, an $N$ x $N$ $M_{ij}$ symmetric interaction matrix, which can be expressed in terms of eigenvalues and eigenvectors as:

$$E = \frac{1}{2}\sum_{i,j=1}^{N} M_{ij} = \frac{1}{2}\sum_{i,j=1}^{N}\sum_{k=1}^{N} \lambda_k(t)W_i^k(t)W_j^k(t)$$

where, $\lambda_k$ is the k[th] eigenvalue, and $W_i^k$ is the i[th] component of the corresponding eigenvector. The eigenvector associated with the most negative eigenvalue contains information about the most and least stabilizing interactions in the system,[18] and we can therefore define an approximate interaction matrix $M_{ij}$ as:

$$\widetilde{M}_{ij} = \lambda_1 v_i^1 v_j^1$$

Under the assumption that residues involved in structural stability are not prone to adaptation, conformation change, and dynamic behavior, we consider the binding interaction with a potential partner as a local phenomenon involving regions not directly dedicated to structural stabilization. From this point of view, we can filter the approximated interaction matrix $M_{ij}$ so that it contains only those pairs of residues that are in geometric proximity in the analyzed structure, resulting in the Matrix of the Local Coupling Energy, or MLCE:

$$MLCE_{ij} = \widetilde{M}_{ij} \otimes c_{ij}$$

Where $c_{ij}$ is the contact matrix of residues, which is worth 1 if the residues $i$ and $j$ are closer than 6 Å (considering $C_\beta$ for proteins), and 0 otherwise, while $\otimes$ is the element-by-element product (i.e., Hadamard product).[39]

Starting with the $MLCE_{ij}$ matrix, we select the residues that have an interaction energy with respect to other residues that belong to the lowest 15% (default value) of the re-ranked pair-energy values. This cutoff value proved to be a key parameter for specificity/sensitivity of the approach. The obtained residues are then fused into patches, which are sets of residues close to each other and constitute the predicted Protein-Protein Binding regions.[39]

To summarize, by analyzing the energetics of residue-pair interactions we can unveil key information about the structural organization and location of the interacting areas of the molecule. The working hypothesis is that specific residue networks may be dedicated to stabilizing folds, while others may be concerned with establishing partner interactions. Weaker pairwise interactions, combined with the localization of residues in continuous areas on the protein surface, highlight substructures that are not internally optimized and are therefore prone to interact with a potential partner.

To this end, we applied MLCE calculations to structural representatives of both the states of NLRP3 (i.e., **ADPstate** and **MCCstate**) which were selected as the most recurrent frames arising from clustering (see description above).

**CNN-ML**

*Preparing DF-images*: The trajectories from the MD-simulations were directly submitted to the DF-matrix calculation using the above reported procedure. Specifically, we extracted the DF each 10ns during the last equilibrated 250ns of each of the four replicas. We ended up with a total number of 208 DF-matrices. We then used an in-house developed *Gnuplot* script (available in Supporting Information) to prepare the images with a dimension of 300X300 pixels using a white-blue-black color palette. Colors tending towards white indicates DF of ~0 Å$^2$, while blue nuances indicate DF of ~10 Å$^2$.

*Preparing the CNN-model*: Image recognition through Convolutional Neural Networks (CNN) was elaborated using a modified version of the readily available VGG19 model, since it demonstrated to be one of the best compromises between computational cost and accuracy and can be directly imported in Python using GPU compiled Tensorflow (TF) (the Python script is available in the Supporting Information).[20]

The architecture of VGG19 model was maintained unaltered, while we modified the dimensions of layers in order to accommodate the 300x300 pixels of the input DF-image. Furthermore, the imported images were again rescaled to the dimension of the VGG19 layers and normalized according to the standard pixel values which can range from 0 to 255. This step aims to exclude possible scaling errors introduced during the *Gnuplot* image preparation from the numerical matrix.

We set the classification mode to 'binary' (class_mode) and the number of samples propagated through the network was set to 32 (batch_size). To provide a measure for goodness of the method we used *ImageNet* weights as widely recognized to be a standard for images classification problems.[21] The sigmoid' activation function $\sigma(z)$ acts in the last layer of our VGG19 modified model by providing the prediction output. We selected this function since it is the standard for binary classifications: given its existence only between 0 and 1, it constitutes the natural choice for binary problems. We compiled the model by using a 'binary cross-entropy' loss function and using the 'Adam' algorithm for the stochastic optimization.[22] Lastly, in order to avoid model overfitting, we introduced an early stopping monitor which stops training the model if the validation loss starts increasing during five consecutive epochs. However, we never experienced strong increases in propagation of loss function to justify the intervention of the monitor. Moreover, the specific placement of the five max pooling layers reduces the computation time and memory usage, by limiting also the probability to get into overfitting issues.

*Training of the model and test with internal data*: To train the model, we selected the DF-images coming from the last equilibrated 250ns of each replica, for a total number of 208 images, which were manually divided between test (20%), train (64%) and validation (16%) sets. Within these sets we operated a manual classification in order to define the two main classes of interest in our model: Active state (**ADPstate**) and Inactive state (**MCCstate**). We trained our model for 20 epochs using the datasets and we obtained complete training and validation in 21 seconds, with an average of 147 ms/step. This result is extremely promising and is mainly due to the TF parallelization of using GPU. We next tested the just trained model with data arriving from the same equilibrated portion of dynamics, but not used during the train and validation steps. We got 100% of accuracy, taking less than 1s to scan all the 42 DF test images. This result was also checked through classification report and confusion matrix analysis, in order to validate the goodness of the predictions (see Supporting Information for more details).

*Test of model with external data*: We submitted to the trained model a new dataset prepared by taking 88 unseen DF matrix images from molecular dynamics replicas of the NLRP3 monomer involved in this study and never used for the previous training of the model. We selected 44 images coming from the Active state (**ADPstate**) and 44 from the Inactive state (**MCCstate**). The manual choice we operated was specifically directed in order to choose within the first non-equilibrated parts of each trajectory. The aim was to prove that – once the model is trained – our method can be extended to other new variants without the need to use long MD simulations. Those images were firstly submitted to the same scaling and normalization steps as performed for the other sets of data (see above). The only difference we introduced was on the number of samples propagated through the network, which was set to 1 (batch_size) since we need to predict each submitted image. Moreover, according with the just printed classification report, we were able to assign the Active state class if the prediction assumes values above 0.5, while if below the Inactive state class was inferred.

Again, the test on external data was extremely fast and only took 9s to complete all the 88 classifications, with a final accuracy of 100% for the Active state and 96% for the Inactive one.

**ASSOCIATED CONTENT**

**Supporting Information.**

Supporting information contains regarding raw data used for the analysis reported in the text: distance distributions, RMSF, all the matrix of distance fluctuations and the point-by-point subtraction of the DF matrix of each replica, contact analysis and temporal contact evolution for both ADP and MCC, full dihedral analysis on R349, F-statistics, Waterswap detailed water contact analysis and local fluctuations. Then, the Neural-Network architecture and the summary of the model with the confusion matrix and Cohen's kappa coefficient.

Supporting Information, AMBER input files, topologies, and restart (rst) files for **APOstate** and **MCCstate** are provided as a GDRIVE directory at the link: https://drive.google.com/drive/folders/1yiejTaKkoExSbOeB0kXwQ_5WZYcnZ2na?usp=sharing

The drive folder is organized in such a way that it is divided into two folders: 'MD' (molecular dynamics) which contains the material for each simulation divided into two folders for the two states including the starting solvated topology and restart files (MCCstate{APOstate}.solv.top{.rst}) together with the input files for all the phases of the molecular dynamics (pre-production and production steps); then the 'CNN-ML' which contains the scripts required for machine learning. Given their large dimensions, full trajectories are available upon request.

**ACKNOWLEDGEMENTS**

The authors thank grant H2020-SGA-FETFLAG-HBP-2019- BRAVE project - FPA No: 650003, Specific Agreement number: 945539, HBP SGA3, for providing support for this research. Funding is also acknowledged through the IMMUNO-HUB, T4-CN-02, project from the Ministero della Salute (Italy). FS and GC thank "Programma di ricerca CN00000013 National Centre for HPC, Big Data and Quantum Computing, finanziato dal Decreto Direttoriale di concessione del finanziamento n.1031 del 17.06.2022." We also thank the Fenix Infrastructure for providing the necessary computing time (Application number 25681), the EOS cluster at University of Pavia and the University of Turin.